**Biological Concepts Instrument (BCI): A diagnostic tool for revealing student thinking**

Michael W. Klymkowsky*✷, Sonia M. Underwood** & R. Kathleen Garvin-Doxas***

*  Molecular, Cellular & Developmental Biology and CU Teach
   University of Colorado, Boulder, Boulder, CO 80309-0347

**  Chemistry, Clemson University, Clemson, SC 29634

*** Center for Computational Language and Education Research,
    University of Colorado, Boulder, Boulder, CO 80309-0594

✷ Corresponding author:  michael.klymkowsky@colorado.edu

**Abstract**:   A key to effective teaching is an awareness and accurate understanding of the thinking and implicit assumptions that students bring to the subject to be learned.  In the absence of extensive Socratic interactions with students, one strategy to assess student thinking involves the use of concept inventories (CIs).  CIs are typically multiple-choice assessments, constructed based on research into student thinking and language, and designed to reveal the presence of common misconceptions and implicit assumptions pertaining to a particular facet of a subject.  Here we describe the open-source Biological Concepts Instrument (BCI), a diagnostic, multiple-choice instrument designed to provide instructors with a preliminary map of a number of basic ideas in molecular level biology.  We describe the strategy behind its design, the research upon which it is based, item construction, and its possible uses as a means to reveal and address persistent and often unrecognized conceptual obstacles.

**Introduction**:

What learning actually is can be the subject of reasonable debate.  From a Socratic perspective (see Cicero, 1989), to have learned something means to understand the assumptions behind it, its implications, and its application (where appropriate) to new situations.  Yet few of us attain this Socratic ideal with our own students, particularly within the high enrollment, introductory sequence courses that are all too common.  In the real world we face a situation analogous to observer bias in scientific studies - we can see what we want (or is in our best interest) to see.  In the context of scientific research, this situation is addressed through double-blind experimental design, and more importantly, independent replication and the building upon previous observations, a process that tests both their validity and the extent of their generality, at least in theory (see Ioannidis 2005; Freedman 2010).

These are ideals that are difficult, impractical, and rather unfortunately rare in the pedagogical arena. In practice, assessment is almost always carried out in an "uncontrolled manner" by the instructor, a person who is not only inextricably involved, both emotionally and professionally, but also potentially impacted by the outcome. It was in this context that instruments designed to examine students' thinking about specific concepts were first introduced (in the post-Socratic era) by Treagust and colleagues (Treagust 1985; Peterson 1986; Treagust 1986; Fetherstonhaugh 1987). An example that has a broad impact on thinking in physics education is the Force Concept Inventory (FCI) (Hestenes and Halloun, 1995; Hestenes et al., 1992). "When he first heard about the FCI, applied physicist Eric Mazur of Harvard University in Cambridge, Massachusetts, assumed that his élite students would perform perfectly well in the traditional lecture setting. So when they received an average FCI score of 70, where 80 is considered a pass, he got "a slap in the face" (Powell, 2003).

At the same time it is worth keeping in mind that many of the original and subsequent claims made for the FCI have not been validated (Huffman and Heller 1995); while a useful diagnostic the FCI like other instruments, provide only a starting point for subsequent instructional interventions. When used thoughtfully, a wide range of instruments (which we will lump together as CIs) can help ground instructors' thinking about their students' understanding and implicit assumptions, which as noted by Huffman & Heller (1995), can often be context specific, for example, "*Even though students understand the Second law with hockey pucks on ice, this does not necessarily mean that they also understand the Second Law with rockets in space.*" "*We use current knowledge structures to help to assimilate new ones more rapidly. To the extent that current beliefs are true, then, we will assimilate further true information more rapidly. However, when the subset of beliefs that the individual is drawing on contains substantial amounts of false information, knowledge projection will delay the assimilation of the correct information*" (Stanovich, 2009). In a classical context, instructor understanding of student thinking would be gained through an on-going Socratic interactions which, when carried out by a perceptive interrogator and receptive students, provides both with important insights that have implications for understanding as well as course and curricular design. Sadly such Socratic interactions are often impractical, infrequent, or absent altogether.

It was in this context that we set out to design a Biological Concepts Instrument (BCI) (Klymkowsky et al., 2003; Klymkowsky et al., 2006). At that time there were a few assessments available, e.g. on diffusion and osmosis (Odom, 1995) and natural selection (Anderson et al., 2002, although see Nehm and Schonfeld, 2010), but neither dealt with the content of our own molecular/cellular biology-based program. In addition, what "counts" as the foundational concepts of biology (analogous to Newton's laws or Maxwell's equations) is much more contentious and poorly defined (Mayr, 1985; Moore, 1993), and questions such as "at what level should the processes of meiosis, or DNA replication, chromatin organization and epistatic regulation, or RNA translation be understood and how many molecular components need to be introduced (or memorized?)" were rarely discussed explicitly. Our approach to conceptual assessment is focused on biology's molecular foundations and based on a research driven analysis of student thinking and language. Although it is unlikely that any multiple choice instrument can provide a truly nuanced view of student thinking, we suggest that the responses to BCI questions can serve to identify topics that instructors may want to examine in terms of their own students' thinking (see Libarkin, 2010.)

**Methods:**

As described previously (Klymkowsky and Garvin-Doxas, 2008), we developed the Biological Concepts Instrument based on student responses to a set of open-ended, short essay questions answered by students at various institutions around the country. A total of 69 such question were asked over the course of the project, a total of 18,286 responses were collected, and samples from each response set were analyzed using the Ed's Tools system (now housed at http://edstools.colorado.edu) (A list of the questions and the number of responses for each is included as supplementary material S1). As described previously, this system enables researchers to collect and analyze student language (access to this "response library" is available to researchers upon request). Many student responses were examined independently by two researchers, who subsequently used them to generate structured student "think-aloud" interview protocols in which a trained interviewer (KGD) questioned students about their thinking on specific topics; ~20 students were interviewed "in depth" where most of these interviews were carried out by KGD or by students under her supervision. Analysis of the interview transcripts led to the construction of multiple-choice questions and their distractors (typically four choices for each question). Most of the questions emerged directly in response to student responses, while others were built more explicitly to probe topics of students' conceptual confusion. A guiding principle in question design was to avoid questions that ask student's to define a term or select an answer solely on terms of vocabulary - this excluded questions such as "What is an allele?" not because they are bad questions *per se*, but because they represent definitions rather than a functional understanding of why alleles matter. Questions were designed so that the responses were nearly equal in length. The face validity of these questions was established through student interviews, which also served to establish that students who picked a particular answer did indeed hold the concept (correct or not) that the answer was designed to represent. Where the two diverged, the question and its distractors were either revised or discarded and replaced, and new interviews were carried out. At the end of this phase of the project, ~80 interviews were performed, and a series of 30 questions that varied in difficulty were selected for inclusion in the first trial version of the instrument. This version was piloted in a number of University biology classes around the country; after that test, one question (question 23) was removed because it had a distinctive, two-tiered structure which tended to confuse students, particularly those responding online.  Content validity was established through discussion of BCI question with disciplinary experts, in some cases, we decided to retain questions that provoked discussion among experts.  All students whose work is discussed here signed informed consent forms (IRB 0304.09 - exempt).

The current 29 item instrument (BCI.1) has been administered across the curriculum at a number of institutions (6), including multiple class sequences throughout the US and Europe.  However for the purposes of this paper, only the results obtained from the introductory and the molecular courses (taken as the fourth course in the sequence) offered in the same department are considered.  These courses are important since they are at the beginning and near the end of a "core" freshman/sophomore molecular curriculum sequence at a large public research university, and so they would provide insights into how instruction might or might not alter student responses to the BCI questions. To determine reliability, the BCI was administered to molecular biology majors in the same courses (Intro and Molecular) three years apart; the first administration is referred to as Intro-1 (255 responses) and Molecular-1 (124 responses), while the last administration is referred to as Intro-2 (254 responses) and Molecular-2 (141 responses).  Both Intro-1 and Molecular-1 were administered the BCI on paper and in class; while Intro-2 and Molecular-2 were administered the BCI online and outside of class; there was no consistent differences between the responses collected on paper or online.  These results were further compared to 85 middle and high school teachers (who answered the BCI online, in response to an email to the National Science Teachers Association biology list service) to examine whether distractor and correct choices differed dramatically between the two groups (students and teachers).

Statistical comparison of the answer choices between different groups and classes used the R statistical package "asbio" under the "KW.multi.comp" topic.  This involved a non-parametric analysis, since the data does not have a normal distribution.  When a Kruskal-Wallis test is performed, a p-value < 0.05 indicates that the groups are statistically significant, but does not allow for the determination of which groups are significantly different; this necessitates the use of a multiple comparison test in addition to the non-parametric ANOVA (Kruskal-Wallis) test.  Adding a multiple comparison test to the Kruskal-Wallis test allows for a more stringent evaluation due to the use of a Bonferroni correction of the α for controlling for Type 1 errors, (that is an erroneous rejection of the null hypothesis).  Only p-values of less than 0.05 were deemed significant.

**Results:**

In contrast to the Force Concept Inventory, which deals with macroscopic processes (that is Newtonian mechanics) and students' working knowledge about the world, the molecular level biological processes that are the focus of the BCI are generally inherently unobservable and therefore student understanding is largely didaskalogenic in origin or arises from the inappropriate application of macroscopic understandings to microscopic processes. As an example, consider the situation of understanding molecular diffusion,

> *"if I stand in a field and hit tennis balls in all directions, balls end up scattered about with a density that declines with distance; if a wall is placed at one point in the field, balls accumulate in front of it. But this it is not how molecules behave; if randomly moving molecules are released from a point source and encounter a barrier, they simply move away from, and around, the obstacle. The ability to sidestep obstacles also explains why freely diffusing molecules will traverse any random maze in not much more time than it takes them to cross the same distance in free space. This result helps us accept what otherwise may seem counterintuitive: that the labyrinth of tortuous intercellular spaces in most tissues poses little impediment to the free diffusion of morphogens."* (from Lander 2007).

In this context it is important to keep in mind that student language does not mean untutored language: students use biological terms and scientific sounding language, even when their grasp of underlying scientific concepts is weak. As an example, during the course of an interview about diffusion:

**Student**: *I have a problem in molecular biology, because I never want to be stupid like "what's diffusion"? because like everyone else ...*

**KGD**: *but diffusion is really hard...*

**Student**: *yeah, and so, I mean, there's a lot of words like that in cellular and molecular, where professors like sometimes just say a word, and I don't know what it means and I'm not going to ask them on it, and I not going to ask you know even a student, because this student will think I am dumb, so I will sometimes go and research but it sometimes I just, I'll start using it without even knowing it.*

Thus the overwhelming majority of responses to an Ed's Tools question on a particular topic are couched in technical language. The following examples, responses to the question "what is diffusion and why does it occur?", typically contain references to gradients, membranes, and other such technical terms, even though the question does not imply such scenarios.

**Student 1:** *"Diffusion is simple transport of material into and out of the cell. More specifically, diffusion is accomplished through the plasma membrane by very small, normally uncharged molecules such as water. Ions and other larger molecules require channels or pumps to cross the membrane and are thus not defined as diffusion. Such non-diffused materials are selectively allowed into the cell, namely, for safety reasons. Foreign substances cannot diffuse across the membrane and the cell's integrity is maintained. Similarly, the cell requires a balance of some materials in and out of the cell to prevent an over-exertion of charge and material gradients that can damage the cell or cause it to explode/implode. Water diffusion is one such balance that is necessary."*

**Student 2:** *"Diffusion is where a small molecule, or uncharged molecule can passively pass through a barrier. When talking about cells, the barrier is usually a membrane such as the plasma membrane. Small, uncharged molecules such as $O_2$, or $CO_2$ can easily pass through a membrane that is permeable to those substances. This passive transport does not require energy, ATP, from the cell. Because this transport does not require energy from the cell, this transport happens often, and the ATP from the cell can be used to transport other substances and molecules into the cell, such as channels that allow ions to pass through."*

**Student 3:** *"diffusion occures (sic) because of differences in concetrations (sic)."*

While the students often wrote in great detail about ideas that are peripheral to the question, what was striking is how rarely students talk about molecular motions as the driving force for all diffusive processes and how diffusion is the result of thermally driven random walks, not gradients or membranes (since there would be no net molecular movement without thermal motion). Similarly, when asked *"How is genetic information stored in an organism and how is it used?"* there were very few students who talk explicitly about the sequence of nucleotides in a DNA molecule and how they specify the sequence of RNAs and polypeptides. Here are some typical responses:

**Student 4:** *"Genetic information is stored in the nucleus of each cell in chromosomes, which are like genetic material "packages." DNA molecules are long strands of nucleic acids, which are like codes for the genetic info., wound (sic) around core histones and twisted and condensed by a seemingly impossible task into the chromosomes. The genetic information is used to create new cells by transcription and translation processes. Transcription is the synthesis of a single strand of RNA from the double-stranded DNA, and the subsequent event of mRNA translation is initiated by ribosomes in the rRNA, and using the tRNA to create the correct amino acid sequence."*

**Student 5:** *"Genetic information is stored in then (sic) DNA of a cell of an organism. The DNA is usually supercoiled in the form of chromatin or a chromosome, depending the organism or its stage of cell divison (sic). The usage of DNA and genes to create and control structure processes is termed expression. Genes are expressed by a system of proteins and translation. Proteins/enzymes act to push forward reactions, which originate from genes formulating these proteins."* [1]

In both responses, students fail to answer the question, they simply restate it: genetic information is in the nucleus or in the DNA. No explanation is given on how it is actually stored, which is what the question asked. Many words are used (nucleus, chromosome, nucleic acids, histones, transcription, translation, ribosomes, rRNA, tRNA, supercoiled, chromatin, and enzymes) but the basic idea, namely that the sequence of nucleotides encodes information, does not appear to be present in most responses.

**BCI Question groups:**

The FCI in particular, and physics in general, benefits from the presence of a working consensus on the key ideas in introductory physics as well as the order in which they should be presented, typically Newtonian mechanics followed in the second semester by electricity and magnetism (not withstanding whether or not this largely historical progression reflects the best, that is most effective, order in which to present physics, or whether it meets the needs of most students taking the course). There is also a distillation of concepts into laws and equations, something not common in most areas of biology. At the same time, agreement on the conceptual foundations of the biological sciences has been rather more difficult to establish (Klymkowsky and Furtak 2009)[2], and there is little evidence for a similar level of agreement as appears to be the case in physics. Our own analysis of the foundations of biological sciences has a decidedly molecular and evolutionary perspective (see Klymkowsky, 2010). In this light, the coverage of the BCI is not meant to be universal or complete, but rather to sample, however sparsely, a number of important foundational concepts.

---

[1] These and other EdsTools responses can be accessed upon request; send requests to michael.klymkowsky@colorado.edu

We have analyzed the BCI questions in 6 broad groups: diffusion and drift (5 questions), energetics and interactions (4 questions), molecular properties and functions (6 questions), genetic behaviors (7 questions), evolutionary processes (5 questions), and experimental design (2 questions). These questions, the offered responses (that is, the correct and distractor answers), and our interpretation of their meanings are supplied in the supplemental materials S2. The questions approach the various underlying themes from different and often superficially distinct directions, and student responses vary with the scenario. It is also the case that a particular question may have relevance to other concept groups. The questions are clustered by general concept rather than by using statistical techniques such as factor analysis. It is important to note that many of these clusters approach a concept from a different angle, for example the energy cluster contains two questions (2 and 3) that require only a basic understanding of how plants and animals use energy (basic – but not necessarily well understood). However the other questions do not explicitly discuss energy, rather students must understand the implicit role of energy in molecular interactions. Clusters such as these would not emerge from factor analysis, and therefore valuable information about differential student understanding of the role of energy in biological systems might not emerge.

**Diffusion and drift group (Questions 1, 5, 25, 29, 30)**: The questions in this cluster are linked by the fact that the underlying processes all involve random events. The five questions in this cluster deal explicitly with molecular level movements (Q1, 25) and genetic drift and related processes (Q5, 29). Question 30 places the two types of processes in juxtaposition. Of these questions, the majority of students answered Q1 above > 50% correct response and that does not change significantly over the course progression. For the other questions, the responses were at or below that expected for guessing (25%). In the case of Q5, students (and teachers) preferred (>50%[3]) "Individuals carrying a1 or a3 were removed by natural selection". In the case of Q25 the preferred response was "its electronegativity would attract it to the ATP synthase," an active (rather than a random) driver, while for Q29 the preferred response was "it produces new combinations of alleles." In all three cases, a response reflecting an active process was preferred to the correct response reflecting the random nature of the underlying process. Particularly in the case of Q29, this is somewhat surprising, since it is the failure to pass all alleles to the next generation that is the primary cause of genetic drift. In the case of Q30, responses tended to be evenly divided between the correct response and "They are not alike. Genetic drift is random; diffusion typically has a direction," which again reveals the idea that diffusion is associated with gradients (see above) and has a preferred direction. We have discussed students' preference for active and directed drivers in what are random processes in more detail elsewhere (Garvin-Doxas & Klymkowsky, 2008). It is interesting to note that there is no statistically significant difference between students' responses for any of these questions except Q25 for one of the upper level classes.

---

[2] A Google search for "conceptual foundations of the biological sciences" produce four hits, while "conceptual foundations of biology" produced ~1100 hits, in contrast a.search for "conceptual foundation of physics" produced ~23,800 hits. Perhaps surprisingly, a search for "conceptual foundations of chemistry" produced 8 hits. Similar results were obtained using the bing search engine ("conceptual foundations of ___" produced 2 hits for chemistry, 19 for biology, and 126 for physics.

[3] Response rates are presented at two significant digits. Graphical data is supplied as supplement 2.

Q1: **Many types of house plants droop when they have not been watered and quickly "straighten up" after watering. The reason that they change shape after watering is because ...**
A. Water reacts with, and stiffens, their cell walls.
B. Water is used to generate energy that moves the plant.
C. Water changes the concentration of salts within the plant.
D. ✔ Water enters and expands their cells.

Q5: **There exists a population in which there are three distinct versions of the gene A (a1, a2, and a3). Originally, each version was present in equal numbers of individuals. Which version of the gene an individual carries has no measurable effect on its reproductive success. As you follow the population over a number of generations, you find that the frequency of a1 and a3 drop to 0%. What is the most likely explanation?**
A. There was an increased rate of mutation in organisms that carry either a1 or a3.
B. Mutations have occurred that changed a1 and a3 into a2.
C. Individuals carrying a1 or a3 were removed by natural selection.
D. ✔ Random variations led to a failure to produce individuals carrying a1 or a3.

Q25: **Imagine an ADP molecule inside a bacterial cell. Which best describes how it would manage to "find" an ATP synthase so that it could become an ATP molecule?**
A. It would follow the hydrogen ion flow.
B. The ATP synthase would grab it.
C. Its electronegativity would attract it to the ATP synthase.

D. It would be actively pumped to the right area.

E. ✔ Random movements would bring it to the ATP synthase.

**Q29: Sexual reproduction leads to genetic drift because ...**

A. there is randomness associated with finding a mate.

B. ✔ not all alleles are passed from parent to offspring.

C. it is associated with an increase in mutation rate.

D. it produces new combinations of alleles.

**Q30: How is genetic drift like molecular diffusion?**

A. Both are the result of directed movements.

B. Both involve passing through a barrier.

C. ✔ Both involve random events without regard to ultimate outcome.

D. They are not alike. Genetic drift is random; diffusion typically has a direction.

**Energetics and interactions (Questions 2, 3, 17, 18):** While energetics and molecular interactions are physicochemical rather than biological concepts, they do play a foundational role (together with molecular properties and functions, see below) in any rigorous understanding of the organization and behavior of biological systems, which are essentially long-lived (> 3 billion years) non-equilibrium systems. The first two questions in this cluster (Q2 and Q3) address the differences between plants and animals (the two most familiar groups of organisms) in terms of how they obtain and use energy. For Q2 the correct response was high (> 65%) for all groups. In contrast (Q3), an apparent understanding that plants capture energy directly through the absorption of photons, while animals must extract energy through the reorganization of molecules, does not rise above 50% in any group. The most attractive alternative response was "Animals use energy to move; plants cannot." The two other questions in this group (Q17 and 18) are more molecular in focus and correct responses fell below that expected for random choice. For Q17, the response "Correctly bound molecules fit perfectly, like puzzle pieces" is the most attractive and reflects a geometric rather than an energetic conceptualization of molecular interactions, a conclusion supported by the analysis of Ed's Tools student responses and student interviews. While common in instructional materials, "geometric complementarity is far from being sufficient to discriminate between native and non-native geometries, except for a very few cases" (Palma et al. 2000, see also Shoichet and Kuntz 1991; Lin 2010). In the case of Q18, "A chemical reaction must change the structure of one of the molecules" is the most attractive distractor, a response that suggests that a specific driver is needed to reverse intermolecular interactions, an assumption that does not apply to a wide range of molecular interactions. Again, there is no statistically significant difference between students' responses for any of these questions. The finding that students and teachers have trouble with these questions is not surprising, given the common preference for directed/active versus random drivers (Garvin-Doxas and Klymkowsky, 2008).

**Q2. In which way are plants and animals different in how they obtain energy?**
A. Animals use ATP; plants do not.
B. ✔ Plants capture energy from sunlight; animals capture chemical energy.
C. Plants store energy in sugar molecules; animals do not.
D. Animals can synthesize sugars from simpler molecules; plants cannot.

**Q3: In which way are plants and animals different in how they use energy?**
A. Plants use energy to build molecules; animals cannot.
B. Animals use energy to break down molecules; plants cannot.
C. Animals use energy to move; plants cannot.
D. ✔ Plants use energy directly, animals must transform it.

**Q17: How does a molecule bind to its correct partner and avoid "incorrect" interactions?**
A. The two molecules send signals to each other.
B. The molecules have sensors that check for "incorrect" bindings.
C. ✔ Correct binding results in lower energy than incorrect binding.
D. Correctly bound molecules fit perfectly, like puzzle pieces.

**Q18: Once two molecules bind to one another, how could they come back apart again?**
A. A chemical reaction must change the structure of one of the molecules.
B. ✔ Collisions with other molecules could knock them apart.
C. The complex will need to be degraded.
D. They would have to bind to yet another molecule.

**Molecular properties and functions group (Questions 10, 11, 13, 19, 20, 27):** This set of questions probes more deeply into the properties of specific biological macromolecules. Because of their central role in all biological systems and their highly stereotyped structure and function, we chose to focus our questions on DNA (Q10, 11, 19) and membranes and lipids (Q13, 20). The final question in this cluster (Q27) could, perhaps, be considered a genetics question, but its aim is to determine whether students can make the leap from the properties and functions of DNA to phenotype. In the case of Q10, 13, and 19 correct responses were at or below that expected for random choices. For Q10, the preferred choices were "The hydrogen bonds that hold it together are very stable and difficult to break" and "The bases always bind to their correct partner."  The first of these is not true, and in fact DNA "breathing" (the opening and closing of the double helix) plays an important role in transcription (Alexandrov et al., 2010; Ambjornsson et al., 2007), while the second, while true, is not the reason that information can be stored in DNA, although it is relevant for DNA replication (see below).  The correct response captures the key structural insight (Watson and Crick, 1953) that because the A=T and G=C base pairs have the same length, the sequence of base pairs does not greatly influence the overall structure of double-stranded DNA.  This makes it possible to store information in base sequence in an unbiased manner. There was no significant difference in the percent correct between these groups.

Correct responses were higher for Q11, which deals explicitly with the mechanism of DNA replication and reflects the insight, derived from the work of Chargaff (Vischer and Chargaff, 1948; Manchester, 2008), of base complementarity, central to Watson and Crick's model.  Question 19 probes an understanding of the properties of catalysts in an unfamiliar context (since the typical presentation of catalysis is in the context of proteins, and on occasion RNAs).  Students overwhelming choose distractor A (*DNA is* "It is stable and does not bind to other molecules") not withstanding the fact that both parts of the response are incorrect, i.e. DNA is not particularly stable (see Discussion) and it readily associates with a wide range of molecules.  In the case of  Q27, which addresses the specific case of haploinsufficiency (a dominant phenotype associated with a null mutation), there is some evidence for progress during the course of instruction.  In all of the student groups, the distractor "If the deleted allele were dominant" was most attractive and can be interpreted as another example of an "active" process, in contrast to the simpler (and correct) idea, "one gene makes less gene product than two genes."

Questions 13 and 20 deal explicitly with the properties of lipids and membranes, universal components of cells.   In the case of Q13 correct responses were around that expected for random choice.  The favorite choice was "whether the molecule is actively repelled by the lipid layer" (another active process).  In the case of Q20 the favored distractor was "Their inability to bond with water molecules" which overlooks the amphipathic (and defining) nature of lipids as a group.

**Q10: What makes DNA a good place to store information?**
A.  The hydrogen bonds that hold it together are very stable and difficult to break
B.  The bases always bind to their correct partner.
C.  ✔ The sequence of bases does not greatly influence the structure of the molecule.
D.  The overall shape of the molecule reflects the information stored in it.

**Q11: What is it about nucleic acids that makes copying genetic information straightforward?**

A. Hydrogen bonds are easily broken.

B. ✔ The binding of bases to one another is specific.

C. The sequence of bases encodes information.

D. The shape of the molecule is determined by the information it contains.

**Q13: When we want to know whether a specific molecule will pass through a biological membrane, we need to consider ...**

A. the specific types of lipids present in the membrane.

B. ✔ the degree to which the molecule is water soluble.

C. whether the molecule is actively repelled by the lipid layer.

D. whether the molecule is harmful to the cell.

**Q20: Lipids can form structures like micelles and bilayers because of ...**

A. their inability to bond with water molecules.

B. their inability to interact with other molecules.

C. their ability to bind specifically to other lipid molecules.

D. ✔ the ability of parts of lipid molecules to interact strongly with water.

**Q19: Why is double-stranded DNA not a good catalyst?**

A. It is stable and does not bind to other molecules.

B. ✔ It isn't very flexible and can't fold into different shapes.

C. It easily binds to other molecules.

D. It is located in the nucleus.

**Q27: Consider a diploid organism that is homozygous for a particular gene. How might the deletion of this gene from one of the two chromosomes produce a phenotype?**

A. If the gene encodes a multifunctional protein.

B. ✔ If one copy of the gene did not produce enough gene product..

C. If the deleted allele were dominant.

D. If the gene encoded a transcription factor.

**Genetic behaviors group (Questions 7, 15, 16, 21, 22, 24, 28):** This set of questions approaches some basic genetic principles associated with phenotype (Q7), genetic linkage (Q15), the definition (Q16) and molecular nature of dominant (Q21) and recessive (Q24) traits, as well as the basic "calculus" of sexual reproduction (Q22) and linkage (Q28). It is worth noting that the second course in the curricular sequence taken by the Intro and Molecular groups deals explicitly with genetics, although the exact content and emphasis of these genetics courses were not necessarily identical.  In the case of Q7 there was a high (65%) level of correct responses.  The correct response to Q22 was again relatively high.  For  Q15, 16, 21, 24, and 28 all distractors appears similar in their attractiveness.

**Q7. If two parents display distinct forms of a trait and all their offspring (of which there are hundreds) display the same new form of the trait, you would be justified in concluding that ...**

A. both parents were heterozygous for the gene that controls the trait.

B. ✔ both parents were homozygous for the gene that controls the trait.

C.  one parent was heterozygous, the other was homozygous for the gene that controls the trait.

D.  a recombination event has occurred in one or both parents.

**Q15: An allele exists that is harmful when either homozygous or heterozygous.  Over the course of a few generations the frequency of this allele increases.  Which is a possible explanation? The allele ...**

A. ✔ is located close to a favorable allele of another gene.

B.  has benefits that cannot be measured in terms of reproductive fitness.

C.  is resistant to change by mutation.

D.  encodes an essential protein.

**Q16: In a diploid organism, what do we mean when we say that a trait is dominant?**

A. It is stronger than a recessive form of the trait.

B. It is due to more, or a more active gene product than is the recessive trait.

C. ✔ The trait associated with the allele is present whenever the allele is present.

D.  The allele associated with the trait inactivates the products of recessive alleles.

**Q21: A mutation leads to a dominant trait; what can you conclude about the mutation's effect?**

A. It results in an overactive gene product.

B.  It results in a normal gene product that accumulates to higher levels than normal.

C. It results in a gene product with a new function.

D. ✔ It depends upon the nature of the gene product and the mutation.

**Q22: How similar is your genetic information to that of your parents?**

A. ✔ For each gene, one of your alleles is from one parent and the other is from the other parent.

B. You have a set of genes similar to those your parents inherited from their parents.

C. You contain the same genetic information as each of your parents, just half as much.

D. Depending on how much crossing over happens, you could have a lot of one parent's genetic

information and little of the other parent's genetic information.

**Q24: A mutation leads to a recessive trait; what can you conclude about the mutation's effect?**

A. It results in a non-functional gene product.

B. It results in a normal gene product that accumulates to lower levels than normal.

C. It results in a gene product with a new function.

D. ✔ It depends upon the nature of the gene product and the mutation.

**Q28: Gene A and gene B are located on the same chromosome. Consider the following cross: AB/ab X ab/ab. Under what conditions would you expect to find 25% of the individuals with an Ab genotype.**

A. It cannot happen because the A and B genes are linked.

B. It will always occur, because of independent assortment.

C. ✔ It will occur only when the genes are far away from one another.

D. It will occur only when the genes are close enough for recombination to occur between them.

**Evolutionary mechanisms group (Questions 4, 6, 12, 14, 26):** This set of questions examines student responses to various evolutionary history/natural selection situations, such as random and catastrophic effects on evolutionary change (Q4), how natural selection produces evolutionary change (Q6), how structures are lost during the course of time (Q12), how mutations might produce novel effects (Q14) and how genetic drift may influence evolutionary history (Q26), a question related to the diffusion/drift question cluster. Q4 produced a high level of correct responses and the most attractive distractor was "New genes are generated" (another type of "active" response to a random event). For Q6 the most attractive distractor was "Producing genes needed for new environments." In the case of Q12, responses were near random and the most popular distractor was "It is no longer actively used". In the case of Q14, there was a bifurcation in distractor choice between "If the mutation inactivated a gene that was harmful" for Intro-1 and Molecular-1 and "If the mutation had no effect on the activity of the gene product" for Intro-2 and Molecular-2. For Q26, which deals with changes in an asexual population, two distractors were particularly attractive, "genetic drift" and "its mutation." Q12 proved most problematic for students, who apparently fail to grasp the "cost-benefit" calculation implicit in evolutionary events. An unexplored area is how students view the development of new traits, and whether they recognize the role of historic contingency, such as illustrated by the development of the ability to utilize citrate in populations of *E. coli* (Blount et al., 2008).

**Q4: How can a catastrophic global event influence evolutionary change?**

A. Undesirable versions of genes are removed.

B. New genes are generated.

C. ✔ Only some species may survive the event.

D. There are short term effects that disappear over time.

**Q6: Natural selection produces evolutionary change by…**

A. ✔ changing the frequency of various versions of genes.

B. reducing the number of new mutations.

C. producing genes needed for new environments.

D. reducing the effects of detrimental versions of genes. …

**Q12: It is often the case that a structure (such as a functional eye) is lost during the course of evolution. This is because ...**

A. It is no longer actively used.

B. Mutations accumulate that disrupt its function.

C. It interferes with other traits and functions.

D. ✔ The cost to maintain it is not justified by the benefits it brings.

**Q14: How might a mutation be creative?**

A. It could not be; all naturally occurring mutations are destructive.

B. If the mutation inactivated a gene that was harmful.

C. ✔ If the mutation altered the gene product's activity.

D. If the mutation had no effect on the activity of the gene product.

**Q26: You follow the frequency of a particular version of a gene in a population of asexual organisms. Over time, you find that this version of the gene disappears from the population. Its disappearance is presumably due to …**

A. genetic drift.

B. ✔ its effects on reproductive success.

C. its mutation.

D. the randomness of survival.

**Experimental design cluster (Questions 8, 9):** We included two general questions aimed at revealing student thinking about experimental design, in particularly whether they could identify a negative control experiment (Q8) and how they might better design the experiment (Q9). For Q8, the most attractive distractor was "It serves as a positive control," which displays a clear misunderstanding of what a positive control is, something that subsequent studies have confirmed (Shi et al., in preparation). For Q9 responses were fairly evenly divided between the correct response and "test only people without opinions, pro or con, about acupuncture," which again suggests confusion as to how control experiments are designed.

**Q8. You are doing experiments to test whether a specific type of acupuncture works. This type of acupuncture holds that specific needle insertion points influence specific parts of the body. As part of your experimental design, you randomize your treatments so that some people get acupuncture needles inserted into the "correct" sites and others into "incorrect" sites. What is the point of inserting needles into incorrect places?**

A. ✔ It serves as a negative control.

B. It serves as a positive control.

C. It controls for whether the person can feel the needle.

D. It controls for whether needles are necessary.

**Q9. As part of your experiments on the scientific validity of this particular type of acupuncture, it would be important to ...**

A. test only people who believe in acupuncture.

B. test only people without opinions, pro or con, about acupuncture.

C. have the study performed by researchers who believe in this form of acupuncture.

D. ✔ determine whether placing needles in different places produces different results.

**Statistical Comparisons:** The answers to individual questions were analyzed (using the statistical methods described in the Methods section) to compare responses from students in the four different courses. When comparing these four groups (Intro-1, Intro-2, Molecular-1, and Molecular-2) we found significant differences between only five out of 29 question (Q 9 & 14 were answered differently between the two Intro groups, while Q 11,16, and 25 were different between the two Molecular groups) (see supplement material S2). It should be noted that these administrations were done three years apart and not with the same sample of students; therefore, any differences could be due to differences in instruction.

      We believe BCI-based comparisons can provide an estimate of whether student understanding changes over time and within the context of this specific curriculum/course sequence. When comparing how the two Intro groups compared to the two Molecular groups, for almost every question except 24 there was no significant difference between the answers for at least three out of the four groups. For the introductory courses there was no significant difference between student answers, and

in fact for almost half of the questions (Q1, 3, 4, 5, 7, 10, 12, 13, 18, 19, 22, 26, 29, 30), there was no significant difference in performance between all four groups, both for questions where students tended to answer correctly (>50%) and to questions where they did not (correct response < 25%). This finding implies that for these questions students at both levels were consistent and retained their understanding or misconceptions in the face of instruction. For the other half of the questions (Q2, 6, 8, 9, 11, 14, 15, 16, 17, 20, 21, 25, 27, and 28) the performance of students in one of the upper level (sophomore/junior) courses was different from the other three. This is probably because of differences in course (and intervening course) content and emphasis, which often vary from year to year and are influenced by the fact that different faculty design and deliver the intervening courses with different emphases. However, for the majority of questions there was no significant difference between introductory and upper level courses. The exception was Q24 where both Molecular courses were found significantly higher than both Intro courses, suggesting, perhaps, an instructional effect.

A second analysis compared middle/high school teacher responses to students. Again, a number of the questions (Q 2↑, 3↓, 5↓, 7↑, 8↑, 9↔, 10↓, 13↓, 18↓, 19↓, and 29↓)( "↑" indicates greater that 50% correct, "↓" indicates less than 50% correct, and "↔" indicates ~ 50% correct for all groups) showed no difference in response implying that for these questions, understanding or misunderstanding are very resilient. However, a second group (Q1, 4, 6, 12, 15, 16, 22, 25, 27, and 30) revealed that at least three of the four student groups were significantly <u>lower</u> than the teacher group suggesting that for these questions the teachers had a better understanding of the concepts then the students. For a third group of questions (Q17, 20, 21, 24, and 28) teachers were significantly higher than the Intro groups but not the Molecular groups. This suggests (but certainly does not prove) that teachers had attained a level equal to, but not significantly better than the upper level student groups. Since most teachers do not come from molecular biology departments, this suggests that 1) molecular biology departments do not do significantly better in these areas than other types of biology departments or ii) in the course of their teaching, teachers gain a similar level of understanding with respect to these questions. There are three remaining questions (Q 11, 14, and 26) that do not fit any of these three main trends. For these questions, the teachers were significantly different from only one of the student groups. What this means, exactly, is not clear and suggests that further studies could be useful.

**Discussion:**

Different assessment instruments have different purposes (see Libarkin, 2010). Standardized tests typically serve as hurdles or sorting systems. The typical classroom exam is often similar (although rather less expensive to generate). Aside from being "too hard" or "too easy", such tests generally provide little in the way of useful (that is explicit) feedback as to students' conceptual strengths or difficulties.  Some conceptual instruments have used the fact that they correlate with averaged final exam scores to justify their validity (see Smith et al., 2008), but given the history of conceptual assessments (see above), this seems problematic.  It is all too easy to teach to a test, once the goals of the test are recognized.  This may be one issue that can arise when instruments like the Force Concept Inventory are used for the pre-/post assessment of student learning.  What would be better would be if there were two distinct but "congruent" versions of the instrument, but this is often impractical.  At the same time, if the focus of an assessment is pedagogically sound, using it to influence course design or teaching strategy would seem quite reasonable as long as independent measures are employed to confirm student understanding.

      The BCI represents a different type of conceptual assessment, quite distinct from standard exams.  Standard tests (summative assessments) are often designed to be efficiently and unambiguously gradable, and so must present questions for which there is only a single "unambiguously true" response that can be recognized by the "best" students (not withstanding the actual complexity of biological systems).  Whether such questions are conceptually informative becomes of secondary import.  In contrast, the point of the BCI and similar instruments is to entice students into revealing their actual thinking about a particular subject.  They are designed to promote learning by drawing instructors' attention to areas where students are having problems, and which standard exams fail to reveal.  There is already an extensive literature on student misconceptions in the biological sciences (Anderson and McKenzie, 2010; Duit, 2009), but this literature is rarely an integral part of the typical instructor's habit of mind.  It is all too easy to become constrained by the logistics of the teaching situation - there is often simply not enough time to consider student thinking on the fly, and to adapt course content.  This is particularly the case for research active professors, obsessed as they are with their other (and often more important, from a career perspective) endeavors, e.g. lab administration, student mentoring, experimental design and interpretation, manuscript and proposal preparation.  We believe that the BCI can provide useful information in this context.  It enables the instructor to focus on key ideas.  This is important, since there can be negative social pressures associated with what are perceived to be "over-simplified" courses, even if the "more rigorous" versions of the course are often not demonstrably more effective at promoting rigorous understanding of key ideas (see Sundberg and Dini, 1993; Sundberg et al., 1994).  The value of assessment instruments like the BCI is that they help instructors, curriculum designers and course analysts appreciate the conceptual obstacles students face and recognize the skills they need.  That said, responses to the BCI cannot be taken as the sole or definitive evidence for a particular level of understanding in students; we see them as a first step in an on-going evaluative process aimed at defining how best to present subject matter and effectively engage students in the often difficult task of mastering that subject matter.  This is a process that involves content revision and various pedagogical strategies, ranging

from the instructor-centric and Socratic to the communal, peer-based and collaborative (Bransford et al., 2001; Klymkowsky, 2007; Mazur, 1997).

It is in this light that we propose that the BCI should be as a diagnostic and formative assessment for identifying potential areas of conceptual confusion and for assaying the effects of course content, sequence, and emphasis. Instruments such as the BCI are based on student language and research into student thinking, and are purposefully written to avoid the clues present in typical instructor generated questions; this can in some situations lead to some measure of ambiguity. This is one reason that they are not intended to serve as a summative assessment. They have a different structure and a different feel from typical exams. Their purpose is to provide insights that are important but often overlooked. The BCI is not a standardized, normed test and it is not intended for pre-post testing, but rather to help instructors discover where students are having difficulty; for this reason we have made BCI questions freely available and encourage others to used them (with appropriate citation).

There are a number of ways that the BCI could be used effectively. One could give it after the core curriculum has been completed, and then use the results to review and, where necessary, redesign course and curricular content and emphasis. Another is to use relevant BCI (or similar) questions formatively, in class; this could involve having students explain their assumptions when picking a correct answer and discussing why the "wrong" responses are wrong, "less right," or dependent upon ancillary assumptions. Ancillary implicit assumptions impact a number of questions where distractors capture plausible specific, but not generally applicable solutions. For example, the two questions on the nature of recessive and dominant alleles are aimed to determine, primarily, whether students understand the complex function that connects genotype, molecular structure, and function with phenotype.

One particularly obvious outcome from our administration of the BCI is that the teacher group did not "Ace" the test (supplemental material S1). This is related to the fact that for a number of questions, responses (including those of teachers) were distributed in a near random manner. Consider, for example, the implications when students (and teachers) hold the idea that DNA is inherently "stable" (Q19). They may well find the processes of mutation and the cellular systems that serve to repair events such as base hydrolysis (depurination and depyrimination), single and double-stranded DNA breaks (Longhese et al., 2006; Seluanov et al., 2010) mysterious, unnecessary, or unbelievable. Understanding DNA repair systems is also the basis for understanding how gene duplication and genomic rearrangements come about, as well as their possible effects on the evolution of new traits, reproductive success and isolation (Raskina et al., 2008).

On a related front, it is also likely that both students and their teachers may not understand the chemistry involved in DNA stability, and why, for example there are enzyme systems dedicated to DNA repair but not protein repair. For example, while the following quote occurs on a biochemistry course web site "DNA is generally quite stable"[4], the truth is that "Although DNA is the carrier of genetic information, it has limited chemical stability" (Lindahl, 1993). Mutations are not the exception, but an integral part of the process of life, and must be actively repaired (Friedberg et al., 2006; Lindahl and Nyberg, 1972). Similarly, the response (to Q19) that suggest that DNA "does not bind to other molecules" reflects an unexpected misconception of the state of DNA within the cell and overlooks the critical (and hopefully upon further reflection obvious) role of DNA-protein interactions in chromosome organization, the epigenetic regulation of gene expression, DNA replication and repair, and the mechanism by which sister chromosomes pair during meiosis, as well as the effects of translocations on meiotic segregation. These are not mysterious processes, but depend upon intermolecular interactions of various types, some of which (for example, the association of histones with DNA) are quite stable. In the same light, an "in class" analysis and discussion of Q19 could be used as a jumping off point for an in-depth, molecular-level exploration of how catalysts in general and biological catalysts in particular work, why polypeptides (in general) and RNAs in specific cases (for example, within the ribosome) act catalytically, as well as the role of enzymatic co-factors in biological systems.

    We welcome instructors to use the BCI as a formative assessment, an assay for what their students bring to their class, and to use BCI and related questions[5] as part of in-class exercises, for example as clicker questions (Caldwell, 2007; Rao and DiCarlo, 2000; Smith et al., 2009), where students are asked to consider and discuss what makes the incorrect choices incorrect. We expect that such an exercise will help instructors (and their students) recognize conceptual areas that require further exploration, and perhaps foster more realistic and effective course and curricular goals and design. Results from the application of the BCI are likely to be particularly critical to identifying insights that may have escaped their students' attention and that are likely to act as serious obstacles to a robust, rigorous, accessible, and transferrable understanding of important ideas. Moreover, when the conceptual foundations of a discipline are considered seriously, it is more likely that teachers (and course/curriculum designers) will be able to recognize the complexity of even "simple" ideas, which often build upon rather sophisticated foundations, and allow sufficient time for students to master them. As one appreciates the complexity of an idea, there may be a realization that "covering" material is not as important as understanding it, which implies the ability to confidently and correctly apply ideas and interpret observations in novel situations. One can hope that the use of a range of conceptual assessments will serve as a brake on syllabus hypertrophy and hyperplasia.

---

[4] http://www.mun.ca/biochem/courses/3107/Topics/DNA_properties.html

[5] with appropriate citation, please

*Klymkowsky et al.,*     20

**Acknowledgements:** *We particularly wish to thank Melanie Cooper for her patient and extensive comments on various versions of the text, many of which we have adopted; we thank Will Leary, Rachel Gheen, Jeffery Ahn, Katherine Henson for their work interviewing students, Jane Jackson for provoking us, and the UC Boulder Discipline-based Education Research (DBER) for an on-going forum to discuss a wide-range of science education issues. This project was support primarily by NSF grant DUE 0405007 and through NSF-Noyce funding for Will Leary and Rachel Gheen.*

**Note:** "✶" indicates that teacher group is significantly different from three of the four
"+" indicates that teacher group is significantly different from both introductory groups
"o" indicates that teacher group is significant different from some of the student groups

**Diffusion and drift group (Questions 1, 5, 25, 29, 30):**

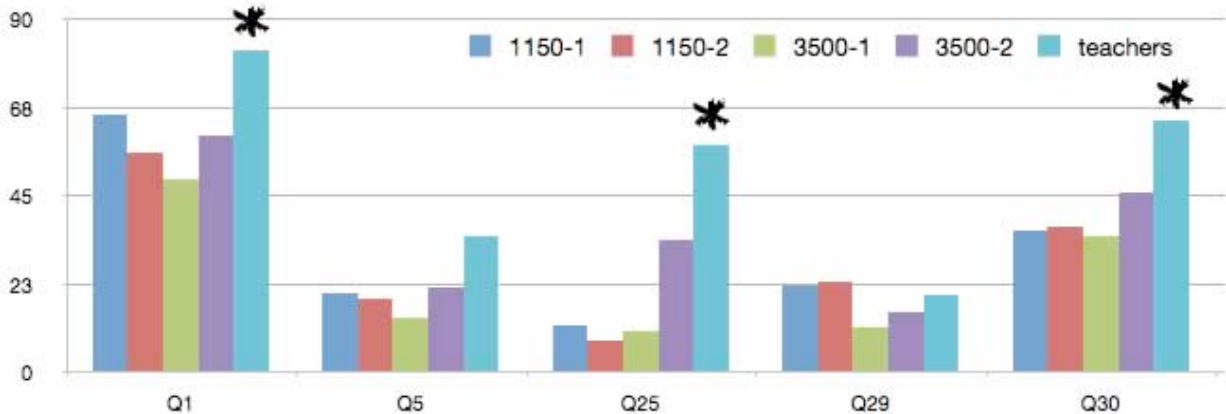

Q1: **Many types of house plants droop when they have not been watered and quickly "straighten up" after watering. The reason that they change shape after watering is because ...**
 A. Water reacts with, and stiffens, their cell walls.
 B. Water is used to generate energy that moves the plant.
 C. Water changes the concentration of salts within the plant.
 D. ✔ Water enters and expands their cells.

Q5: **There exists a population in which there are three distinct versions of the gene A (a1, a2, and a3). Originally, each version was present in equal numbers of individuals. Which version of the gene an individual carries has no measurable effect on its reproductive success. As you follow the population over a number of generations, you find that the frequency of a1 and a3 drop to 0%. What is the most likely explanation?**
A. There was an increased rate of mutation in organisms that carry either a1 or a3.
B. Mutations have occurred that changed a1 and a3 into a2.
C. Individuals carrying a1 or a3 were removed by natural selection.
D. ✔ Random variations led to a failure to produce individuals carrying a1 or a3.

Q25: **Imagine an ADP molecule inside a bacterial cell. Which best describes how it would manage to "find" an ATP synthase so that it could become an ATP molecule?**
A.   It  would follow the hydrogen ion flow.
B.  The ATP synthase would grab it.
C.  Its electronegativity would attract it to the ATP synthase.
D.  It would be actively pumped to the right area.
E.  ✔   Random movements would bring it to the ATP synthase.

Q29: **Sexual reproduction leads to genetic drift because ...**
A.  there is randomness associated with finding a mate.
B.  ✔ not all alleles are passed from parent to offspring.
C.  it is associated with an increase in mutation rate.
D.  it produces new combinations of alleles.

**Q30: How is genetic drift like molecular diffusion?**
A. Both are the result of directed movements.
B. Both involve passing through a barrier.
C. ✔ Both involve random events without regard to ultimate outcome.
D. They are not alike.  Genetic drift is random; diffusion typically has a direction.

**Energetics and interactions (Questions 2, 3, 17, 18):**

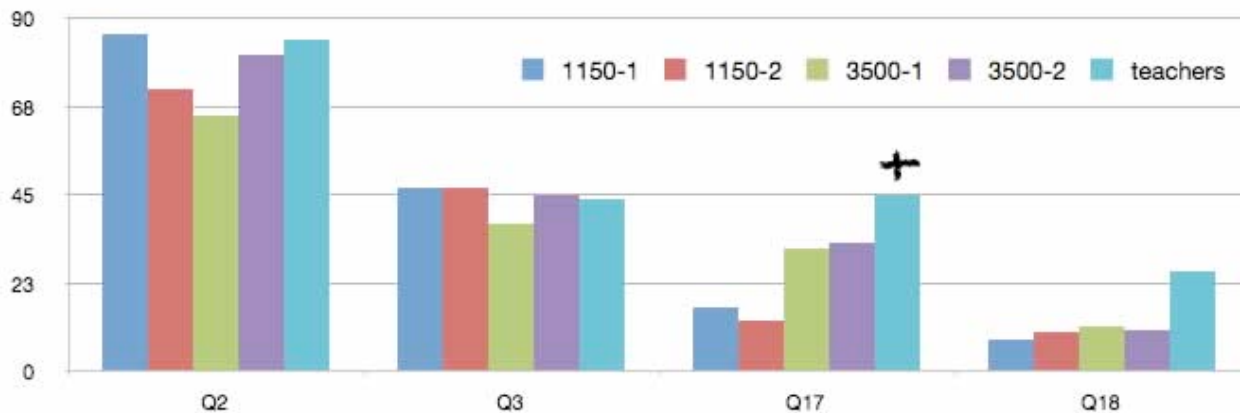

**Q2. In which way are plants and animals different in how they obtain energy?**
A. Animals use ATP; plants do not.
B. ✔ Plants capture energy from sunlight; animals capture chemical energy.
C.  Plants store energy in sugar molecules; animals do not.
D. Animals can synthesize sugars from simpler molecules; plants cannot.

**Q3: In which way are plants and animals different in how they use energy?**
A. Plants use energy to build molecules; animals cannot.
B.  Animals use energy to break down molecules; plants cannot.
C.  Animals use energy to move; plants cannot.
D. ✔ Plants use energy directly, animals must transform it.

**Q17: How does a molecule bind to its correct partner and avoid "incorrect" interactions?**
A. The two molecules send signals to each other.
B. The molecules have sensors that check for "incorrect" bindings.
C. ✔ Correct binding results in lower energy than incorrect binding.
D. Correctly bound molecules fit perfectly, like puzzle pieces.

**Q18: Once two molecules bind to one another, how could they come back apart again?**
A.  A chemical reaction must change the structure of one of the molecules.
B. ✔ Collisions with other molecules could knock them apart.
C.   The complex will need to be degraded.
D.  They would have to bind to yet another molecule.

**Molecular properties and functions group (Q10, 11, 13, 19, 20, 27):**

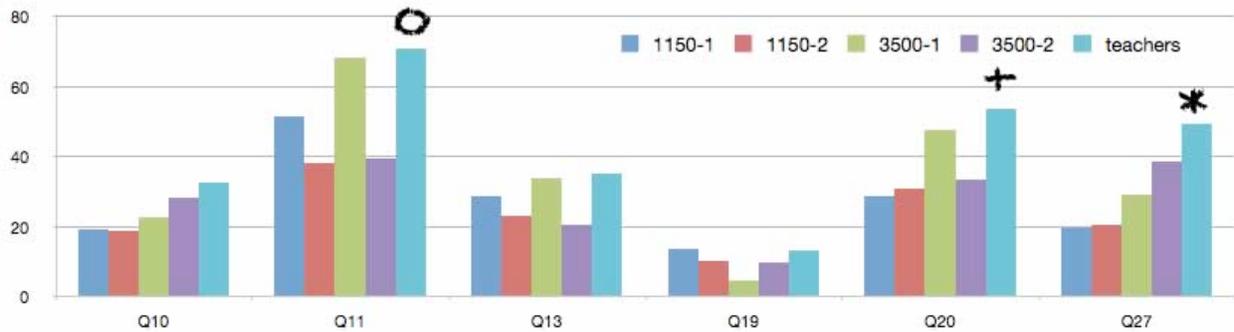

**Q10: What makes DNA a good place to store information?**
A.   The hydrogen bonds that hold it together are very stable and difficult to break
B.   The bases always bind to their correct partner.
C. ✔ The sequence of bases does not greatly influence the structure of the molecule.
D.   The overall shape of the molecule reflects the information stored in it.

**Q11: What is it about nucleic acids that makes copying genetic information straightforward?**
A.  Hydrogen bonds are easily broken.
B. ✔ The binding of bases to one another is specific.
C. The sequence of bases encodes information.
D.  The shape of the molecule is determined by the information it contains.

**Q13: When we want to know whether a specific molecule will pass through a biological membrane, we need to consider ...**
A. the specific types of lipids present in the membrane.
B. ✔ the degree to which the molecule is water soluble.
C. whether the molecule is actively repelled by the lipid layer.
D. whether the molecule is harmful to the cell.

**Q19: Why is double-stranded DNA not a good catalyst?**
A.  It is stable and does not bind to other molecules.
B. ✔  It isn't very flexible and can't fold into different shapes.
C.  It easily binds to other molecules.
D.  It is located in the nucleus.

**Q20: Lipids can form structures like micelles and bilayers because of ...**
A. their inability to bond with water molecules.
B. their inability to interact with other molecules.
C. their ability to bind specifically to other lipid molecules.
D. ✔ the ability of parts of lipid molecules to interact strongly with water.

**Q27: Consider a diploid organism that is homozygous for a particular gene.  How might the deletion of this gene from one of the two chromosomes produce a phenotype?**
A. If the gene encodes a multifunctional protein.
B. ✔ If one copy of the gene did not produce enough gene product..
C. If the deleted allele were dominant.
D. If the gene encoded a transcription factor.

**Genetic behaviors group (Q7, 15, 16, 21, 22, 24, 28):**

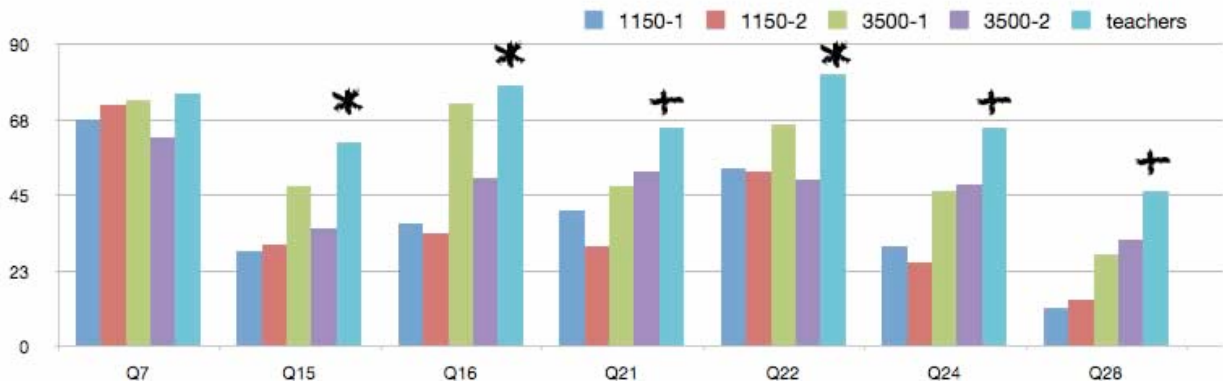

**Q7. If two parents display distinct forms of a trait and all their offspring (of which there are hundreds) display the same new form of the trait, you would be justified in concluding that ...**
A. both parents were heterozygous for the gene that controls the trait.
B. ✔ both parents were homozygous for the gene that controls the trait.
C. one parent was heterozygous, the other was homozygous for the gene that controls the trait.
D. a recombination event has occurred in one or both parents.

**Q15: An allele exists that is harmful when either homozygous or heterozygous. Over the course of a few generations the frequency of this allele increases. Which is a possible explanation? The allele ...**
A. ✔ is located close to a favorable allele of another gene.
B. has benefits that cannot be measured in terms of reproductive fitness.
C. is resistant to change by mutation.
D. encodes an essential protein.

**Q16: In a diploid organism, what do we mean when we say that a trait is dominant?**
A. It is stronger than a recessive form of the trait.
B. It is due to more, or a more active gene product than is the recessive trait.
C. ✔ The trait associated with the allele is present whenever the allele is present.
D. The allele associated with the trait inactivates the products of recessive alleles.

**Q21: A mutation leads to a dominant trait; what can you conclude about the mutation's effect?**
A. It results in an overactive gene product.
B. It results in a normal gene product that accumulates to higher levels than normal.
C. It results in a gene product with a new function.
D. ✔ It depends upon the nature of the gene product and the mutation.

**Q22: How similar is your genetic information to that of your parents?**
A. ✔ For each gene, one of your alleles is from one parent and the other is from the other parent.
B. You have a set of genes similar to those your parents inherited from their parents.
C. You contain the same genetic information as each of your parents, just half as much.
D. Depending on how much crossing over happens, you could have a lot of one parent's genetic information and little of the other parent's genetic information.

**Q24: A mutation leads to a recessive trait; what can you conclude about the mutation's effect?**
A. It results in a non-functional gene product.
B. It results in a normal gene product that accumulates to lower levels than normal.
C. It results in a gene product with a new function.
D. ✔ It depends upon the nature of the gene product and the mutation.

**Q28: Gene A and gene B are located on the same chromosome. Consider the following cross: AB/ab X ab/ab. Under what conditions would you expect to find 25% of the individuals with an Ab genotype.**
A. It cannot happen because the A and B genes are linked.
B. It will always occur, because of independent assortment.
C. ✔ It will occur only when the genes are far away from one another.
D. It will occur only when the genes are close enough for recombination to occur between them.

**Evolutionary mechanisms group (Q4, 6, 12, 14, 26):**

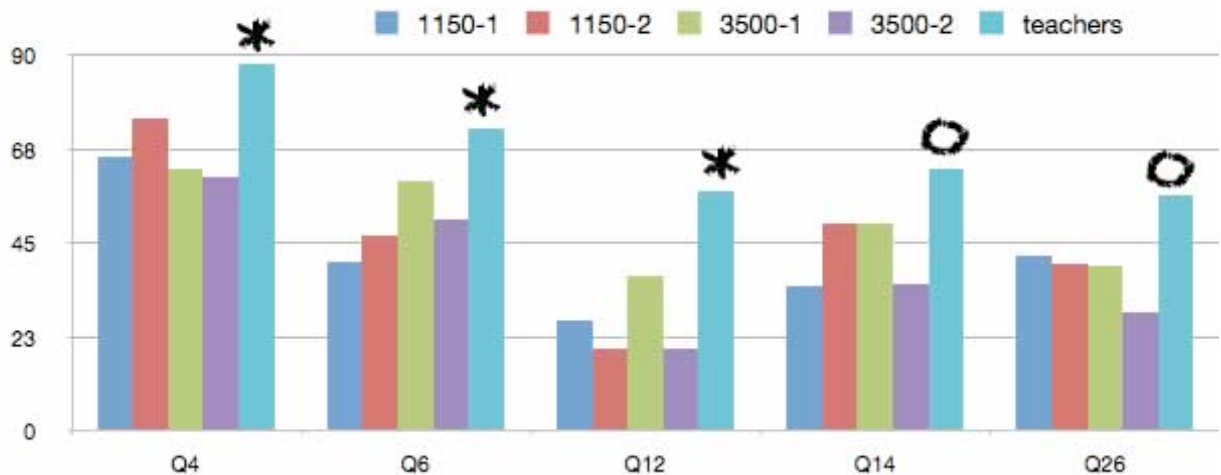

**Q4: How can a catastrophic global event influence evolutionary change?**
A. Undesirable versions of genes are removed.
B. New genes are generated.
C. ✔ Only some species may survive the event.
D. There are short term effects that disappear over time.

**Q6: Natural selection produces evolutionary change by…**
A. ✔ changing the frequency of various versions of genes.
B. reducing the number of new mutations.
C. producing genes needed for new environments.
D. reducing the effects of detrimental versions of genes. …

**Q12: It is often the case that a structure (such as a functional eye) is lost during the course of evolution. This is because ..**.
A. It is no longer actively used.
B. Mutations accumulate that disrupt its function.
C. It interferes with other traits and functions.
D. ✔ The cost to maintain it is not justified by the benefits it brings.

**Q14: How might a mutation be creative?**
A. It could not be; all naturally occurring mutations are destructive.
B. If the mutation inactivated a gene that was harmful.
C. ✔ If the mutation altered the gene product's activity.
D. If the mutation had no effect on the activity of the gene product.

**Q26: You follow the frequency of a particular version of a gene in a population of asexual organisms. Over time, you find that this version of the gene disappears from the population. Its disappearance is presumably due to …**
A. genetic drift.
B. ✔ its effects on reproductive success.
C. its mutation.
D. the randomness of survival.

**Experimental design cluster (Q8, 9):**

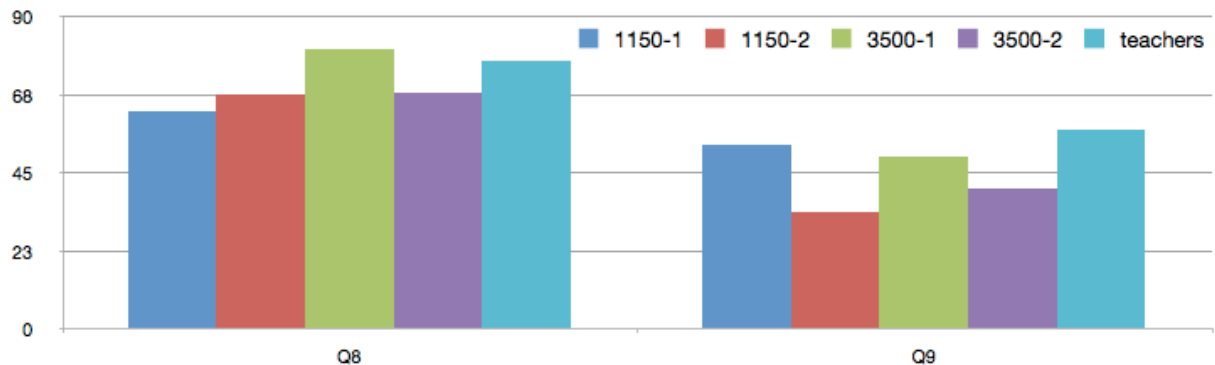

**Q8. You are doing experiments to test whether a specific type of acupuncture works. This type of acupuncture holds that specific needle insertion points influence specific parts of the body. As part of your experimental design, you randomize your treatments so that some people get acupuncture needles inserted into the "correct" sites and others into "incorrect" sites. What is the point of inserting needles into incorrect places?**
A. ✔ It serves as a negative control.
B. It serves as a positive control.
C. It controls for whether the person can feel the needle.
D. It controls for whether needles are necessary.

**Q9. As part of your experiments on the scientific validity of this particular type of acupuncture, it would be important to ...**
A. test only people who believe in acupuncture.
B. test only people without opinions, pro or con, about acupuncture.
C. have the study performed by researchers who believe in this form of acupuncture.
D. ✔ determine whether placing needles in different places produces different results.